\documentclass[12pt]{article}
\usepackage{epsfig}

\parskip=\medskipamount	% space between paragraphs (TeX)
\renewcommand{\thesection}{\arabic{section}}

\catcode`@=11

\@addtoreset{equation}{section}
\@addtoreset{equation}{subsection}
\def\theequation{\ifnum\value{section}=0 \arabic{equation}\ignorespaces
\else \ifnum\value{section}=-1 A.\arabic{equation}\ignorespaces
\else \ifnum\value{subsection}=0 \thesection.\arabic{equation}\ignorespaces
\else \thesection.\arabic{subsection}.\arabic{equation}\ignorespaces
                             \fi
                        \fi
                   \fi}

{\catcode`\'=\active \def'{{}^\bgroup\prim@s}}

\catcode`@=12

\newcommand{\bq}{\begin{equation}}
\newcommand{\be}{\begin{equation}} 
\newcommand{\fq}{\end{equation}}
\newcommand{\ee}{\end{equation}}
\newcommand{\bea}{\begin{eqnarray}}
\newcommand{\eea}{\end{eqnarray}}

\def\tM{{\tilde M}_2}

\def\vw{\vec w}
\def\bop#1{\setbox0=\hbox{$#1M$}\mkern1.5mu
	\vbox{\hrule height0pt depth.04\ht0
	\hbox{\vrule width.04\ht0 height.9\ht0 \kern.9\ht0
	\vrule width.04\ht0}\hrule height.04\ht0}\mkern1.5mu}
   
\begin{document} 

\thispagestyle{empty}

\begin{flushright}
\begin{tabular}{l}
CTP-MIT-3073\\ 
 
\end{tabular}
\end{flushright}

\vskip.3in
\begin{center}
{\Large\bf String Junctions and 
\vskip .2cm
Non-simply Connected Gauge Groups}
\vskip.3in

\vskip .2in
{\bf Zachary Guralnik}
\\[5mm]
{\em Center for Theoretical Physics \\
Massachusetts Institute of Technology\\
Cambridge MA, 02139}\\
{email: zack@mitlns.mit.edu}

\vskip.5in minus.2in

{\bf Abstract}

\end{center}

Relations between the global structure 
of the gauge group in elliptic F-theory compactifications, 
fractional null string junctions, and
the Mordell-Weil lattice of rational 
sections are discussed.  We extend results in the literature, 
which pertain primarily to rational elliptic surfaces and  
obtain $\pi^1( {\tilde G})$ where ${\tilde G}$ is the 
semi-simple part of the gauge group.  
We show how to obtain the full global structure 
of the gauge group,  including all $U(1)$ factors.
Our methods are not restricted to rational elliptic surfaces.
We also consider elliptic K3's and K3-fibered Calabi-Yau
three-folds.

\setcounter{page}{0}  
\newpage 
\setcounter{footnote}{0}

%%%%%%%%%%%%%%%%%%%%%%%%%%%%%%%%%%%%%%%%%%%%%%%%%%%%%%%%%%%%%%%%%%%%%%%%%%%%%%%

\section{Introduction}

The theory of string junctions provides a useful tool 
for  studying aspects of IIB 
string theory,  or equivalently F-theory compactifications 
on elliptically fibered manifolds $X$. 
In this paper we shall use the string junction
technology developed in \cite{DZ, DW,  DHIZ1, DHIZ2,
DHIZ3} to elaborate on the work of \cite{AM} and \cite{FY}
concerning the global properties of the gauge group and
the Mordell-Weil lattice of the the elliptic fibration.

In particular we will show how to  
determine the global structure of the gauge group in F-theory 
compactifications on elliptic K3 manifolds, and propose a method
to  obtain the global structure for K3-fibered 
Calabi-Yau three-folds.  
The gauge groups which can arise in this context are subgroups of 
a single infinite dimensional Lie group \cite{DHIZ1, DHIZ2},  
which is a broken symmetry of the F-theory compactification.
This infinite dimensional group is  
simply connected, having a weight lattice equal to the root lattice.  
We shall show that information about 
the global structure of the subgroups is simply encoded in the 
lattice of null string junctions.

Locally, the gauge group $G$ is 
a product of simple Lie groups and some $U(1)$ factors.  
We will write the semi-simple part of the gauge group (containing no
$U(1)$ factors) as $\tilde G$.
The discussion of the global structure of the gauge group in 
\cite{AM} pertains to $\pi^1({\tilde G})$. 
The authors of \cite{AM} considered compactification on 
an elliptic K3 in the limit of a stable degeneration to a pair 
of intersecting rational elliptic surfaces $R_1$ and $R_2$. 
In this limit,  the states carrying gauge charges
correspond to certain elements of $H_2({\cal R}_i,Z)$,  which may be
represented by string junctions in the base of ${\cal R}_i$.
The map from the elements of $H_2({\cal R}_i,Z)$  to the weight lattice
was found to have a non-trivial cokernal.
The missing representations of the algebra 
were found to imply the  gauge group  
${\tilde G} = {\tilde G}_1 \times {\tilde G}_2$,
with $\pi^1({\tilde G}_i) = T(\Phi_i)$,  where 
$T(\Phi)$ is the torsion part of the Mordell-Weil
lattice of ${\cal R}_i$.
The Mordell-Weil lattice $\Phi$ (see \cite{M-W}) is the lattice of
rational sections of the elliptic surface,  which are closely related to the  
poles of the Seiberg-Witten differential \cite{NTY, SW}.  $T(\Phi)$  
consists of sections $S$ with the property that $nS = 0$,  the zero section,
for some integer n.  

The Mordell-Weil lattice has been completely tabulated
for all rational elliptic surfaces \cite{OS}.   
In \cite{FY} it was observed by 
direct comparison that  $T(\Phi)$ for a rational elliptic surface
is identical to a lattice generated by ``improper'' null
junctions on the $P^1$ base.
These null junctions carry fractional (p,q) charges,  and are not 
realizable as
membranes upon lifting to F-theory.  Their relation to geometrical 
objects,  i.e. sections of the 
rational elliptic surface, is not at all manifest.  
By combining the observations of \cite{FY} with
the results of \cite{AM},  one concludes that these null junctions 
must somehow generate $\pi^1({\tilde G}_i)$.  

We will give a direct argument showing why these null junctions generate
$\pi^1({\tilde G}_i)$.  To do so we will make use of an isomorphism
between the center of the universal cover of $\tilde G$ and the quotient
of the string junction lattice,  which includes ``improper'' fractional
junctions,  by the lattice of proper junctions.
There is a map from the fractional null junctions to the subgroup of the 
center of the universal cover which acts trivially on 
all states in the spectrum.
This map gives not just $\pi^1({\tilde G})$,  but the complete 
global structure of
${\tilde G}$.  Furthermore, this map is not be restricted
to rational elliptic surfaces.  
We shall also consider elliptic K3's and K3 fibered Calabi-Yau three-folds,
taking into account states which decouple in the limit in which a K3 becomes
a pair of intersecting rational elliptic surfaces.  
These states are missing in 
the analysis of \cite{AM},  as they correspond to infinitely massive strings
stretching between the $P^1$ base of each rational elliptic surface. 
In the dual heterotic description on $T^2$,  these states are 
strings wrapping cycles of
the $T^2$,  the radii of which go to infinity as the K3 degenerates 
to a pair of rational elliptic surfaces.
The presence of these states,  which 
do not decouple for a generic K3, modifies the global 
structure of the gauge group computed in \cite{AM}.  This modification can
be determined from the lattice of null junctions in the base of the K3.  

The null junctions give the global structure of the semi-simple part 
of the gauge group, ${\tilde G}$.  One can easily extend 
the analysis to compute the full global structure 
including all the $U(1)$ components.  Generically,  the $U(1)$ 
components of the gauge group do not enter as 
global factors.  The methods we use to compute $\pi^1({\tilde G})$ 
can be easily extended to compute the correlation between $U(1)$ charges 
and charges under the center of ${\tilde G}$.

Some new subtleties appear when analyzing the spectrum of F-theory compactified
on elliptically fibered Calabi-Yau three-folds.  These subtleties arise due to 
singularities of the discriminant curve. The gauge group and hypermultiplet spectrum
of such compactifications have been studied by a variety of 
authors \cite{W, IMS, BIKMSV, KV, AKM, AM}.
The simplest case to analyze involves K3 fibered Calabi-Yau three-folds,  
taking again the limit
in which generic K3 fibers become pairs of intersecting rational 
elliptic surfaces.
In \cite{AM}, it was suggested that the relation between 
the torsion component of the 
Mordell-Weil lattice and $\pi^1(\tilde G)$ might persist for such a Calabi-Yau.
No attempt has yet been made to address this question in general.  The 
Mordell-Weil lattice is more difficult to obtain on an elliptic three-fold.  Furthermore, unlike
the rational elliptic surface,  there is no known theorem relating 
$\pi^1({\tilde G})$ to the 
torsion component of the Mordell-Weil lattice,  although in some 
instances (see \cite{AM})
this relation still holds.  
Nonetheless, we shall argue that one can still use string junction 
technology to compute 
the global properties of the gauge group. It 
is possible to define a string junction lattice on an elliptically 
fibered Calabi-Yau three-fold (see \cite{US}),  
and to associate hypermultiplets with elements of this lattice.  
Unlike a K3, some elements of the string junction lattice 
of the three-fold $X$ are not contained in $H_2(X,Z)$.   
Currently, we do not know if 
$\pi^1({\tilde G})$ is isomorphic to $T(\Phi)$ in all instances.

The organization of this paper is as follows. 
Section 2 is a brief review of string junction lattices in the context of F-theory on an 
elliptic surface. In section 3,  we show why fractional null junctions 
generate $\pi^1(\tilde G)$.
In section 4,  we show how to compute the global structure of the gauge group
including all the $U(1)$ factors.
In section 5,  we review some facts about gauge groups and hypermultiplets in 
elliptically fibered Calabi-Yau three-folds, and propose
a method to obtain the global structure of the gauge group using string
junctions.

\section{Review of string junctions}

The string junction lattice associated with F-theory compactified on 
an elliptic 
surface was defined and studied in \cite{DZ, DW,  DHIZ1, DHIZ2, DHIZ3}.  
We give a very brief review of this work below.
The reader familiar with this technology may wish to skip to the next section.

F-theory on an elliptic surface $X$ describes IIB string theory 
on $P^1$,  with 
7-branes at points in the $P^1$. 
In the absence of other branes (such as a three-brane) the string junction 
lattice consists of the equivalence classes of string junctions having 
endpoints on the 7-branes.  In the F-theory description,  an individual 7-brane 
corresponds to a point in the base (a discriminant locus) over which there is a 
degenerate elliptic fiber 
of Kodaira type
 $I_1$ \cite{K}, having a unique vanishing cycle. 
To define the junction lattice,  one initially considers all seven branes to be 
separated in $P^1$. An example of a string junction is illustrated in figure.1.
Each segment of the junction is an oriented string with charges $(p,q)$,
which are conserved at branching points.   
These  charges correspond to one cycles in the elliptic fiber over each 
segment, and the 
the strings junctions lift to membranes in F-theory.  

Individual 7-branes,  labelled by an index $i$,  carry 
relatively prime charges $(p_i, q_i)$ indicating the vanishing cycle of
the $i$'th $I_1$ fiber.  A string ending on a 7-brane must carry the same 
charges. 
Unless a three-brane probe is present,  
the membrane associated with a string junction does not have a boundary,  
and represents an  element of $H_2(X,Z)$.

\begin{figure}
\begin{center}
\epsfig{file=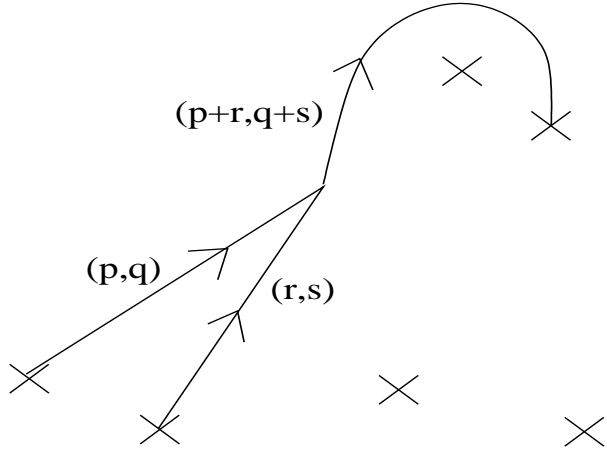,height=6cm,width=8cm}
\end{center}
\caption{ An example of a string junctions in $P^1$.  The crosses
indicate 7-branes. } 
\end{figure}

The vanishing cycle of the $i$'th $I_1$ fiber is the eigenvector of the 
monodromy matrix $K_i$ associated with that
fiber;
\be
K_i = 
\pmatrix{ 
1+p_iq_i & -p^2_i \cr
q_i^2 & 1-p_iq_i \cr}
\ee
To determine the equivalence classes of string junctions on $P^1$, 
it is convenient to move the 7-branes onto the real axis with branch cuts extending 
vertically downward.  A segment of a string junction having charge  
$(p,q)$ on the left side of the branch cut of the i'th 7-brane has charge $K_i(p,q)$ 
on the right side of the branch cut.
Such a segment can be pulled across the 7-brane so that it
no longer crosses the branch cut.  However,  because of an effect 
dual to a Hanany-Witten transition \cite{HW},
an additional segment appears stretching  between the original segment 
and the 7-brane, as illustrated in figure 2.
The charges of the new segment are determined by charge conservation to be 
$(p^{\prime},q^{\prime}) = K_i(p,q) - (p,q)$, which is proportional to
the vanishing cycle $(p_i, q_i)$. 
The two junctions related by pulling this 
segment across the 7-brane 
are said to be in the same equivalence class.  Thus, the 
equivalence classes of junctions can be 
determined by considering only junctions which lie entirely above
the real axis,  as in figure 2(b).  The equivalence classes are then
labelled by vectors ${\vec J} = (N_1,N_2,\cdots N_n)$,
where the integers $N_i$ indicate the number and orientation
of strings ending on the $i$'th 7-brane.
The charges of the string segment ending on the i'th 7-brane 
are $N_i(p_i,q_i)$.  
For junctions which lift to a membrane without a boundary, the total $(p,q)$ charge vanishes;  
$\sum_i N_i(p_i,q_i) = (0,0)$.  For the moment,  we will not make this restriction
and consider membranes with a $(p,q)$ cycle as a boundary.
Later,  to obtain states in F-theory in the absence of three-branes,  we will 
glue such membranes together to get one without a boundary. 

\begin{figure}
\begin{center}
\epsfig{file=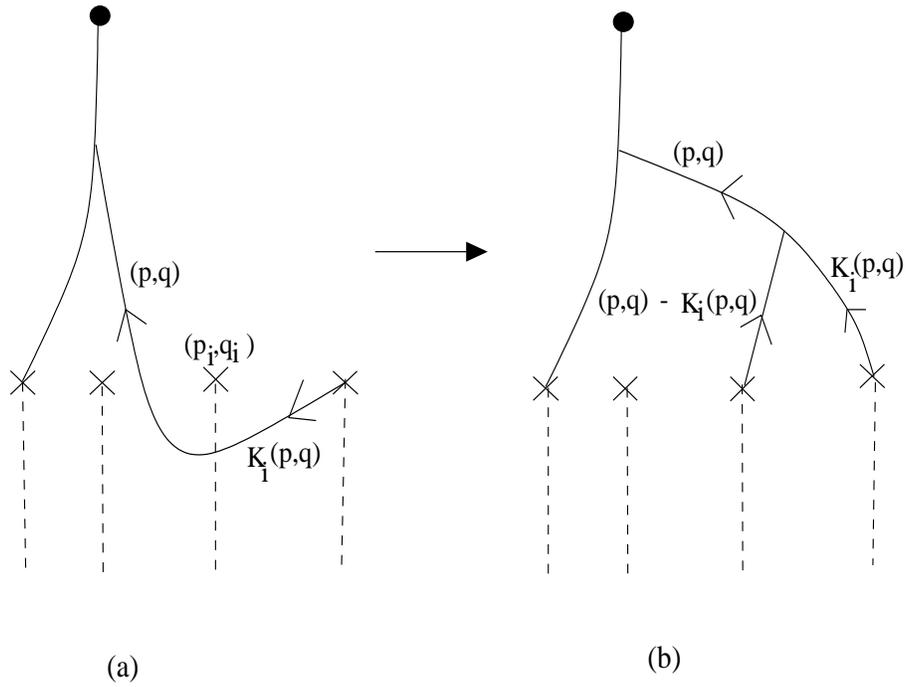,height=9cm,width=12cm}
\end{center}
\caption{ Two equivalent string junctions related by a Hanany-Witten 
transition.
Figure 2(b) is a canonical representation of the equivalence class,  in which
the junction lies entirely above the real axis. 
The dashed lines are branch cuts going down from each 7-brane.     }
\end{figure}

The basis in which an equivalence class ${\vec J}$ is labelled by the integers
$N_i$ is not the most useful one.  
By allowing certain unphysical junctions with fractional $N_i$ charges, one can find
another basis in which there is a a manifest correspondence between elements of the 
junction lattice and weights of a Lie Algebra.   The physical junctions then   
span a sub-lattice of the full string junction lattice.  
Consider the sub-lattice of junctions which end on a certain subset of 
the 7-branes,  and which have
a total asymptotic charge $(p,q) =\sum_i N_i(p_i,q_i)$.  Let us assume that this subset of
7-branes is collapsible.  In the F-theory description,  
this means that the corresponding
$I_1$ fibers can collide to form a more singular Kodaira fiber.   The 
more singular Kodaira 
fibers have an A-D-E classification,  indicating the enhanced gauge symmetry algebra 
of the configuration of coincident 7-branes.  The junction lattice is initially defined before
collapsing the 7-branes. In order to map elements of the junction lattice to weights of the algebra,  
one requires a quadratic form on the junction lattice.  This form was defined in 
\cite{DZ} and, for a $K3$, coincides 
with the intersection form on $H_2(X,Z)$ when considering junctions with vanishing $(p,q)$.
Linearly independent junctions $\vec{\alpha}_k$ with vanishing $(p,q)$ and self-intersection number 
$-2$ are associated with the simple roots of the (simply laced) Lie algebra.  The self intersection matrix of 
these junctions is minus the Cartan matrix; $(\vec{\alpha}_k, \vec{\alpha}_l) = - C_{kl}$.
When 7-branes collide,  these junctions have zero length,  giving rise to massless vector multiplets
and an enhanced gauge symmetry.  As a simple example, figure 3 illustrates the simple 
root junction for two $I_1$ fibers which collide to give an $I_2$ (or $A_1$)  singularity.
In this case, the resulting gauge symmetry algebra  is  that of $SU(2)$.

\begin{figure}
\begin{center}
\epsfig{file=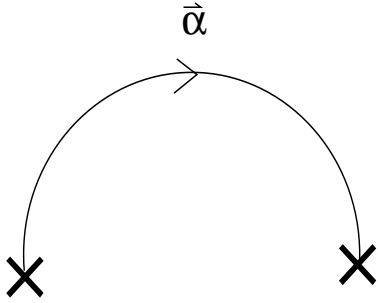,height=4cm,width=5cm}
\end{center}
\caption{ The simple root junction of an $I_2$ Kodaira fiber, 
which has been split into two $I_1$ fibers,  corresponding to 
7-branes with the same charge. The associated algebra is    
$SU(2)$.}
\end{figure}

One can define a set of fundamental weight junctions ${\vec w}_i$ with 
vanishing $(p,q)$ satisfying  
$( {\vec w}^i, {\vec{\alpha}}_j ) = -\delta_i^j$.  These junctions are 
generically ``improper'' 
meaning that may have fractional $N_i$ charges.
To get a complete basis,  one defines another pair of junctions
${\vec w}_p$ and ${\vec w}_q$,  known as ``extended weights,'' 
which are orthogonal to the weight junctions ${\vec w}_i$
and which carry charges $(p,q)=(1,0)$ and $(0,1)$ respectively.
The junctions $\vw_p$ and $\vw_q$ are also generically improper.
Any proper junction $\vec J$ can be written as a linear combination of 
the improper basis junctions.
\be
{\vec J} = p{\vec w}_p + q{\vec w}_q + a^i{\vec w}_i.
\ee
where the integers $a^i$ are the Dynkin labels.  
The integer asymptotic charges $p$ and $q$
are related to $U(1)$ charges and are called ``extended Dynkin labels''.

Multiplets under various representations of the algebra 
may be found by identifying highest weights, however 
it will not be necessary for us to do this,  since we are seeking only
the global structure of the gauge group.

For F-theory on an elliptic surface,  or IIB theory on $P^1$, 
there will generically be several sets 
of coincident
7-branes labelled by an index $m$.  These coincident sets  
give rise to a gauge group which is locally the product of 
A-D-E groups for each $m$ and some $U(1)$
factors.  We will refer to the semi-simple part of the gauge group as ${\tilde G}$,  and
the entire group, including the $U(1)$ factors, as $G$.
A general string junction can be written as
\be
{\vec J} = \sum_m p_m {\vec w}_{p,m} + q_{m}{\vec w}_{q,m} 
+ a^i_{m}{\vec w}_{i,m}. 
\ee

Since we assume the absence of three-branes,  all strings must end on 7-branes and the
total asymptotic charge vanishes,
\be
\sum_{m} p_{m} = \sum_{m} q_{m} = 0.
\label{vanish}
\ee
Here again $p_{m}$, $q_{m}$,  $a^i_{m}$ are 
taken to be integers. 
For a physical (or proper) string junction,  the charges $N_i$ 
must also be integer.
Combining this constraint with the vanishing of the total asymptotic charge
restricts the representations which can appear in the spectrum.  
We will see later that this restriction is consistent with a 
particular non-simply connected gauge group.

Note that one could consider all the 7-branes collectively, instead of grouping
them according to the unbroken gauge symmetry,  and write 
\be
{\vec J} = \sum_i a_i {\vec w}_i.
\ee
Here one should view the ${\vec w}_i$ as the weights of an infinite dimensional 
Lie algebra \cite{DHIZ1,DHIZ2},  of which only a finite dimensional 
sub-algebra is realized as a gauge symmetry.  In this case the weight lattice is
the same as the root lattice, and the weights ${\vec w}_i$ are proper.
It follows that there is no constraint on the representations which may appear,
and the infinite dimensional Lie group is simply connected. 

\begin{figure}
\begin{center}
\epsfig{file=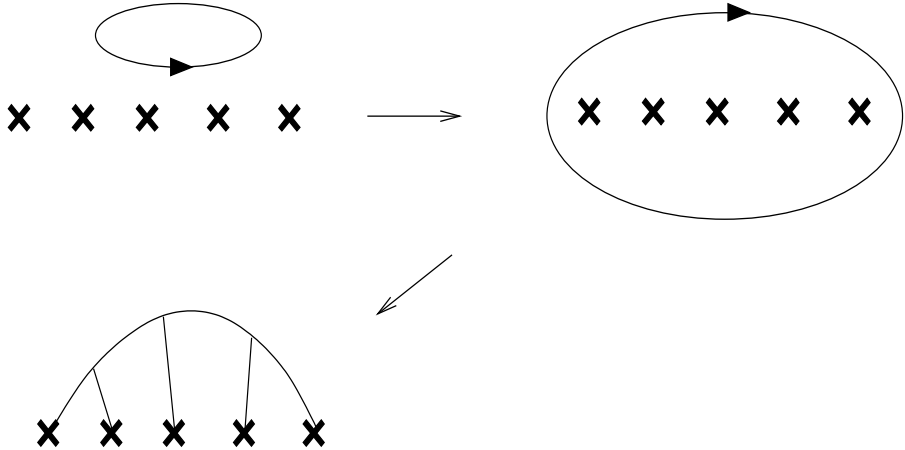,height=6cm,width=12cm}
\end{center}
\caption{A proper null junction,  which is related by Hanany Witten transitions to 
a contractable closed string loop.  The loop surrounds all the 7-branes,  
around which the total $SL(2,Z)$ monodromy is $1$. }
\end{figure}

A subset of the the integers $p_{m}$ and $q_{m}$ are charges for the 
$U(1)$ factors in the gauge symmetry group.  There are actually four fewer 
conserved $U(1)$ 
charges than the number of indices $m$ would indicate.
Two of the charges are not independent,  due to (\ref{vanish}).
Two more of the charges are not conserved, 
due to an equivalence relation in which one adds 
``proper null junctions'' to $\vec J$.  
 
The null junctions are of particular interest in this paper.  
These junctions have vanishing intersection with any other junction.
The {\it proper} null junctions are topologically trivial,  meaning that 
they are related
by Hanany-Witten transitions to a contractable closed string loop.  
If the base manifold
on which the closed string junctions live is a $P^1$ (as for an elliptic 
K3, or a 
rational elliptic surface),  then any contractable closed string loop can 
be viewed 
as a loop surrounding all of the 7-branes,  as in figure 4.  
The total monodromy upon encircling all the 7-branes is the identity.
There are $12$ 7-branes in the base of a rational elliptic surface,  and
$24$ 7-branes in the base of an elliptic K3. 
Via Hanany-Witten transitions, such a loop is equivalent to a junction
lying above the real axis, with endpoints on the 7-branes. This junction
carries no symmetry charges,  and thus has the form 
\be
{\vec N} = \sum_m \left( P_m{\vec w}_{p,m} + Q_m{\vec w}_{q,m} \right).
\ee
with $\sum_m P_m = \sum_m Q_m = 0$.
The $(p,q)$ charge on the
upper part of a string loop surrounding all the 7-branes is arbitrary,  
so the lattice of null junctions is two dimensional.

We shall also have reason to consider ``improper null junctions'' which
are fractions of the proper null junctions, with integer $P_m$ and $Q_m$  
but fractional $N_i$ charges.  We will eventually find that the quotient 
of the lattice of all null junctions (proper and improper) by the 
lattice of proper null junctions
is isomorphic to
 $\pi^1({\tilde G})$,  where ${\tilde G}$ is the semi-simple part
of the gauge group.

\section{Non-simply connected gauge groups for elliptic surfaces}

The gauge symmetry algebra of  F-theory compactified on $K3$ can be read directly from
the A-D-E type of the Kodaira fibers in the K3.  However 
determining the gauge group $G$
takes more work.  This problem was discussed 
in \cite{AM},   where $\pi^1({\tilde G})$ 
was obtained by studying the spectrum for compactifications on  K3's which 
have degenerated to pairs of intersecting rational elliptic surfaces.
We shall address this problem using a method which is not restricted
to rational elliptic surfaces, and which yields the full global 
structure of the gauge group.
In doing so, we will explain the observations of \cite{FY} 
relating improper null junctions to the torsion component of the   
Mordell-Weil lattice of a rational elliptic surface.

If $\pi^1(G)$ is non-trivial,  hypermultiplets in 
certain representations of the algebra can 
not appear in the spectrum.  States charged under $G$ are realized 
as string junctions with vanishing 
asymptotic $(p,q)$.  
The constraint of vanishing $(p,q)$  
together with the requirement that physical string junctions
have integer charges is sufficient to determine the global structure 
of the gauge group.

Consider an elliptic fibration over $P^1$,
with Kodaira fibers $K_m$ (coincident 7-branes) at points $m$ in $P^1$.  
The gauge symmetry algebra ${\cal G}$ is the direct sum of the A-D-E 
algebras of each Kodaira fiber,
${\cal G} = \sum_m {\cal G}_m$.   
There is a sub-lattice of string junctions
associated with each $m$,  comprised of junctions 
having endpoints only at the point $m$.
A junction associated with the m'th point may be written as 
\be
{\vec J}_m  = p_m {\vec w}_{p,m} + q_m{\vec w}_{q,m} + a_{i,m}{\vec w}_{i,m}.
\ee
To obtain physical states (in the absence of three-branes) these 
junctions must be 
combined to give one without any asymptotic charge;
\be
{\vec J} = \sum_m {\vec J}_m,
\ee
with $\sum_m p_m = \sum_m q_m =0$.
One must also impose the constraint that each ${\vec J}_m$ is
a proper junction. Of course not all such states will be BPS, or stable,
but must nevertheless be considered to obtain the global properties of the
gauge group.

Each ${\vec J}_m$ belongs to some representation 
$R_m$ \footnote{The exact representation can not be determined without
knowing the highest weight.  However this is irrelevant to our purposes
since the behavior under the center of the universal cover does 
not depend on the highest weight.}
of the universal covering group $G_m$ associated with the algebra 
${\cal G}_m$.  $G_m$ has a center $C_m$.
We will shortly see that when ${\vec J}_m$ is a proper junction,  
the action of the center $C_m$ in the representation $R_m$ 
is determined entirely by the asymptotic charges
$p_m$ and $q_m$.  In the 
absence of constraints on $p_m$
and $q_m$, the gauge group (up to $U(1)$ factors) would be a direct 
product of the universal covers $G_m$. However there are elements
in the center of $\prod G_m$ which act trivially in all representations
containing proper junctions with $\sum_m p_m = \sum_m q_m =0$. 
This is what gives rise to a non-simply connected gauge group.

Before making the above statements more precise, we shall need some 
simple facts about group theory.  
The center $C$ of the universal cover of a simple Lie group 
has the structure $\prod_l Z_l$,  and a corresponding lattice ${\cal L}_C$.
${\cal L}_C$ is given by the quotient of the 
weight lattice by the root lattice,
\be
{\cal L}_C = {\cal L}_{\vec w}/{\cal L}_{ {\vec \alpha}}
\ee
This quotient gives a natural map from weights to the center lattice.
This map determines how a particular representation
transforms under the center.      
For instance for $SU(2)$ ${\cal L}_C$ is a $Z_2$ lattice comprised of $0$ 
and $1$  (corresponding to center elements $1$ and $-1$ of $SU(2)$).  
The fundamental 
weight ${\vec w}$ maps to $1$.  
A generic weight ${\vec W} = n{\vec w}$ maps to $n \ mod \ 2$.  
Integer spin representations, with weights of even $n$, transform 
trivially under the center,  while
half integer spin representations, with odd $n$, do not. 
In general the root vectors,  or  weights in 
the adjoint representation,  map to the trivial element of the center. 

More generally, suppose a weight vector maps to an element $c$ in the 
center lattice ${\cal L}_C$.  Let $\Psi$ be an element of a representation
containing that weight vector.  Then
under the action of an element of the center represented
by $c^{\prime}$ in ${\cal L}_C$,  $\Psi$ transforms as follows
\be
\Psi \rightarrow \exp(i {\bf c}^{\prime} \cdot {\bf c}) 
\Psi 
\ee
where the 
dot product is taken with respect to an appropriate metric on ${\cal L}_C$.  
For $SU(2)$, where ${\bf c}$ and ${\bf c}^{\prime}$
take values $0$ or $1$ (mod 2), the above phase factor is $\exp(i \pi {\bf c} 
{\bf c}^{\prime})$.

\subsection{Junctions and representations of the center; an $I_3$ example.}

For the moment,  let us just consider string junctions 
with endpoints on a single set of coincident 7-branes,  with no 
requirement that the asymptotic $p$ and $q$ vanish.
For the sake of clarity,  we will work in the context of a particular example in 
which ${\vec J}$ are junctions ending on the locus of an $I_3$ fiber,  which is associated with
an $SU(3)$ algebra.  
The $I_3$ fiber may be split into three $I_1$ fibers, 
each with vanishing cycle $[p,q] = [1,0]$.  
Equivalence classes of junctions are labelled by three charges,
${\vec J} = (N_1, N_2, N_3)$.  In this basis, the intersection form is 
$g_{ij} = -\delta_{ij}$.
The simple root junctions are 
\bea
{\vec \alpha}_1 = (1,-1,0) \nonumber \\
{\vec \alpha}_2 = (0,1,-1).
\eea
The fundamental weight junctions are 
\bea
{\vec w}_1 = (2/3, -1/3. -1/3) \nonumber \\
{\vec w}_2 = (1/3, 1/3, -2/3)
\eea
There is a single extended weight in this case, given by
\be
{\vec w}_p = (1/3, 1/3, 1/3).
\ee
We define the full junction lattice to consist of 
all junctions of the form ${\vec J} = p{\vec w}_p + a_1 {\vec w}_1 + a_2 {\vec w}_2$ with 
integer $p, a_1$ and $a_2$.
In terms of the $N_i$ charges,  one has
\bea
p = N_1 + N_2 +N_3 \nonumber \\
a_1 = N_1 - N_2 \nonumber \\ 
a_2 = N_2 -N_3.
\eea
The full junction lattice contains a sub-lattice of proper junctions with integer
$N_i$.  

Let us restrict ourselves for the moment to the case $p=0$.
For $p=0$, the quotient of the string junction lattice by the lattice of proper junctions is
the same as the quotient of the weight lattice by the root lattice. In this case, 
this quotient  gives $Z_3$.  We will define a linear map, $M_1$,  which maps weight vectors
to the $Z_3$ lattice according to this quotient;
\bea
M_1({\vec w}_1) =  2  \ mod \ 3 \nonumber \\
M_1({\vec w}_2) = 1  \ mod \ 3.
\label{marp}
\eea
Strictly speaking,  the quotient gives a map is defined only
up the automorphism of $Z_3$ which exchanges
$2  \ mod \ 3$ with $1  \ mod \ 3$.  
In (\ref{marp}) we have made a particular choice.
For $p=0$, the proper junctions are in the root lattice and are mapped to 
zero by $M_1$. 
We extend the action of the map $M_1$ to junctions with non-vanishing 
$p$ as follows;
\be
M_1({\vec w}_p) = 0 \ mod \ 3.
\ee

The map $M_1$ determines the behavior of junctions under the center of $SU(3)$.
Let us represent an element of the center by an element $\bf c$  in the 
$Z_3$ lattice,
the action of this element of the center on a component 
$\Psi_{\vec J}$ of a representation
containing ${\vec J}$
is given by  
\be
\Psi_{\vec J} \rightarrow \exp \left( i {\bf c} \cdot M_1({\vec J}) \right) 
\Psi_{\vec J} = 
\exp \left( \frac{2\pi i}{3} {\bf c} M_1({\vec J})\right) \Psi_{\vec J}.
\ee
Note that we have made a choice of metric with which to take the dot product. 
This metric is well defined only up to an automorphism of $Z_3$.  The element
of the center which ${\bf c}$ represents depends on this choice.

We now observe that any combination of weight vectors $a_i{\vec w}_i$, can be made into
a proper junction by adding integer multiples of ${\vec w}_p$. 
For instance, ${\vec w}_p + {\vec w}_1$ is a proper junction, 
whereas ${\vec w}_1$ is
not.  This permits us to define 
another linear map $M_2$ from junctions to the center lattice,  
such that $M_2$ maps
all the proper junctions to zero. $M_2$ acts on the weight vectors the same way as 
$M_1$;
\bea
M_2( {\vec w}_1) = M_1( {\vec w}_1) = 2  \ mod \ 3 \nonumber \\
M_2( {\vec w}_2) = M_1( {\vec w}_2) = 1  \ mod \ 3.
\eea
but acts non-trivially on ${\vec w}_p$;
\be
M_2( {\vec w}_p) = 1  \ mod \ 3.
\ee
One can easily verify that if a junction ${\vec J}_{\cal P}$ is proper, then
\be
M_2 ({\vec J}_{\cal P}   ) = 0.
\label{zmap}
\ee
Due to (\ref{zmap}), the behavior of any proper junction 
under the center, which depends on 
its map under $M_1$,  is
determined entirely by its asymptotic charge $p$;
\bea
M_1({\vec J}_{\cal P} ) = M_1 (p{\vec w}_p + a_1 {\vec w}_1 + a_2 {\vec w}_2)= 
\nonumber \\
M_1( a_1 {\vec w}_1 + a_2 {\vec w}_2) = M_2 ( a_1 {\vec w}_1 + a_2 {\vec w}_2)= 
\nonumber \\
-M_2(p{\vec w}_p) = - p \ mod \ 3.
\eea

It will also prove useful to define linear maps ${\tilde M}_1$ and
${\tilde M}_2$ of junctions to the cover $Z$  of $Z_3$;
\bea
{\tilde M}_1({\vec w}_1) =  2 \nonumber \\
{\tilde M}_1({\vec w}_2) = 1 \nonumber \\
{\tilde M}_1({\vec w}_p) =0
\eea
and
\bea
{\tilde M}_2( {\vec w}_1) = 2 \nonumber \\
{\tilde M}_2( {\vec w}_2) = 1 \nonumber \\
{\tilde M}_2( {\vec w}_p) = 1.
\label{tilmaps}
\eea 
These maps have the property that division is well defined,  in the sense that 
\be
\frac{1}{l}{\tilde M}({\vec J})={\tilde M}(\frac{\vec J}{l})
\label{divprop}
\ee
whenever $l$ is integer and ${\vec J}/l$ is a junction with integer $p$, $q$ and $a_i$.
We shall eventually use null junctions  obtained from
fractions of proper null junctions to represent the elements in
the center of the universal cover of $\tilde G$ which act trivially
on all states in the spectrum.

\subsection{Junctions and representations of the center for a general A-D-E singularity}

In general,  the following is true for any collection of 7-branes 
associated with an A-D-E algebra.

\noindent{\it The quotient of the lattice of junctions ${\cal J}$
by the lattice of proper junctions ${\cal J}_{\cal P}$ is the 
lattice ${\cal L}_C$  associated with
the center of the universal cover of ${\tilde G}$.}
\be
{\cal L}_C = {\cal J}/{\cal J}_{\cal P}    
\ee 
There exist two linear maps from the junction lattice ${\cal J}$ to the
center lattice ${\cal L}_C$,   $M_1$ and $M_2$,  with the following properties.
$M_2$ maps junctions by quotienting by 
the proper junctions,
\be
M_2 ({\vec J}) = {\vec J} \ mod \ {\cal J}_{\cal P}.
\ee  
while $M_1$ maps junctions to the center lattice according to
\be
M_1({\vec J}) = M_1 ( p {\vec w}_p +  q {\vec w}_q + a_i{\vec w}_i ) = M_2 (a_i {\vec w}_i )
\ee

The behavior of junctions under the center of the universal cover is 
determined by the map $M_1$.
The action of an an element of the center represented by ${\bf c} \in 
{\cal L}_{C}$ in a representation containing the junction ${\vec J}$
is given by 
\be
\Psi_{\vec J} \rightarrow \exp(i{\bf c} \cdot M_1({\vec J})) \Psi_{\vec J}
\ee
where the dot product is with respect to an appropriate metric on
the center lattice.
The element of the center which ${\bf c}$ represents is fixed only after
choosing the maps $M_1,M_2$  and the metric,  both of which are defined only up
to an automorphism of the center.   
  
Since $M_2$ maps proper junctions ${\vec J}_{\cal P}$ to zero, 
the action of $M_1$ on proper junctions
is determined entirely by the asymptotic charges $p$ and $q$.
\be
M_1 ({\vec J}_{\cal P}) = M_1 (p{\vec w}_p + q{\vec w}_q + a_i {\vec w}_i ) = 
\nonumber \\ 
M_2 (a_i {\vec w}_i ) =  -M_2( p{\vec w}_p + q{\vec w}_q )     
\ee
For the A-D-E groups,  the center of the universal cover is either of the form
$Z_N \times Z_M$ or $Z_N$,  which is consistent with the fact that there 
at most two asymptotic charges, $p$ and $q$.  
One can also define maps ${\tilde M}_1$ and ${\tilde M}_2$ 
of junctions to $Z \times Z$ or just $Z$ as in (\ref{tilmaps}).
    
Note that the center of the universal cover is isomorphic to
the quotient of the lattice of junctions of the 
form $P{\vec w}_p + Q{\vec w}_q$ by its proper sub-lattice.  
The reader may explicitly verify this by inspecting 
table 1(a) and table 1(b). 
We will later find it useful to represent elements in the center of
the universal cover ${\bf c}$ by junctions of this form;
\be
{\bf c} = M_2({\vec J}_c) = M_2(P{\vec w}_p + Q{\vec w}_q)
\ee
The action of ${\bf c}$ in a representation containing proper junctions 
${\vec J}_{\cal P} = p{\vec w}_p + q{\vec w}_q + \cdots$
can be written as
\be
\Psi_{{\vec J}_{\cal P}} \rightarrow \exp(-i{\tilde M}_2 (P{\vec w}_p + Q{\vec w}_q) 
\cdot {\tilde M}_2(p{\vec w}_p + q{\vec w}_q)) \Psi_{{\vec J}_{\cal P}}.   
\ee

\vspace{40pt}

\begin{tabular}{|l|l|l|l|}
\hline
Kodaira - ADE type & universal cover & center & decomposition \\ 
\hline
$III - A_1$ & $SU(2)$ & $Z_2$ & $AAC$ \\
\hline
$IV - A_2$ & $SU(3)$ & $Z_3$ & $AAAC$ \\
\hline
%%%$I_0^* - D_4$ & $Spin(8)$ & $Z_2 \times Z_2$ & $AAAABC$ \\
%%%\hline
$I_{2n}^* - D_{4+2n}$ & $Spin(8+4n)$ & $Z_2\times Z_2$ & $A^{4+2n}BC$ \\
\hline
$I_{2n+1} - D_{4+2n+1}$ & $Spin(10+4n)$ & $Z_4$ & $A^{5+2n}BC$ \\
\hline
$IV^* - E_6$ & $E_6$ & $Z_3$ & $AAAAABCC$ \\
\hline
$III^* - E_7$ & $E_7$ & $Z_2$ & $AAAAAABCC$ \\
\hline
$II^* - E_8$ & $E_8$ & $I$ & $AAAAAAABCC$ \\
\hline
$I_n - A_{n-1}$ & $SU(n)$ & $Z_n$ & $A^n$ \\
\hline
\end{tabular}
\vspace{10pt}

\noindent Table 1(a).  We list the Kodaira/ADE singularities,  
the center of the universal cover, 
and a decomposition into $I_1$ fibers (or 7-branes) with monodromy 
matrices of type $A, B$ or $C$ arranged from left to right on the 
real axis.  These matrices are (up to an overall $SL(2,Z)$ conjugation),

\be
A = \pmatrix{1 &-1\cr 0&1\cr}, \ \ B = \pmatrix{5& -16\cr 1&-3\cr}, \ \ 
C = \pmatrix{3 & -4\cr 1& -1\cr}
\ee

\vspace{30pt}

\begin{tabular}{|l|l|l|} 
\hline
fiber &  ${\vec w}_p$  & ${\vec w}_q$ \\
\hline
$III$ & $(\frac{1}{2}, \frac{1}{2}, 0)$ & $(\frac{1}{2},\frac{1}{2},-1)$ \\ 
\hline
$IV$ & $(\frac{1}{3},\frac{1}{3}, \frac{1}{3}, 0)$ & $(\frac{1}{3},\frac{1}{3}, \frac{1}{3}, -1)$ \\
\hline
$I_0^*$ & 
$(0,0,0,0,\frac{1}{2},\frac{1}{2})$&$ (\frac{1}{2}, \frac{1}{2}, \frac{1}{2}, \frac{1}{2}, -\frac{3}{2}, -\frac{1}{2})$ \\
\hline
$IV^*$ & $(-\frac{1}{3}, -\frac{1}{3},-\frac{1}{3},-\frac{1}{3},-\frac{1}{3},\frac{4}{3},\frac{2}{3},\frac{2}{3})$&$
(1,1,1,1,1,-3,-1,-1)$ \\
\hline
$III^*$ & $(-\frac{1}{2},-\frac{1}{2},-\frac{1}{2},-\frac{1}{2},-\frac{1}{2},-\frac{1}{2},2,1,1)$&$
(\frac{3}{2},\frac{3}{2},\frac{3}{2},\frac{3}{2},\frac{3}{2},\frac{3}{2},-5,-2,-2)$ \\
\hline
$II^*$ & $(-1,-1,-1,-1,-1,-1,-1,4,2,2)$&$ (3,3,3,3,3,3,3,-11,-5,-5)$ \\
\hline
$I_n$ & $(\frac{1}{n},\frac{1}{n}, \cdots )$ & \\
\hline
\end{tabular}
\vspace{10pt}

\noindent Table 1(b). This table lists some Kodaira fibers and the extended weights 
${\vec w}_p$ and ${\vec w}_q$ written in the $N_i$ basis.

\vspace{30pt}

\subsection{$\pi^1(G)$ for elliptic surfaces.}

To construct physical string junctions arising for compactifications 
of F-theory on an elliptic surface,  one 
must glue together junctions with endpoints on different collections 
of coincident 7-branes 
so that the total $(p,q)$ charge vanishes.
%%% and the junction lifts to a representative of 
%%%$H_2(X,Z)$.  
Junctions then have the form 
\be
{\vec J} = \sum_m {\vec J}_m = \sum_m (p_m {\vec w}_{p,m}+ 
q_m {\vec w}_{q,m} + a_{i,m}{\vec w}_{i,m}).
\ee
with $\sum_m p_m = \sum_m q_m = 0$.
The universal cover of the semi-simple part of the gauge group 
is given by $\prod_m G_m$,  with center
\be
{\cal C} = \prod_m {\cal C}_m.
\ee
We denote an element in the lattice ${\cal L}_C$ associated with 
${\cal C}$ by $\bf c$, 
and an element in the sub-lattice associated with  ${\cal C}_m$ by ${\bf c}_m$.
In general ${\cal C}_m$ is a lattice of the form 
$Z_{(N_m)} \times Z_{(M_m)}$ or $Z_{(N_m)}$.  
The maps $M_1$ and $M_2$ defined in the previous section are now defined for each 
index $m$,  giving 
maps ${\vec J} \rightarrow {\cal L}_C$.
The action of an element of ${\cal C}$ in  
a representation containing the junction ${\vec J}$ is
given by 
\be
\Psi_{\vec J} \rightarrow \exp(i {\bf c} \cdot M_1({\vec J})) \Psi_{\vec J} = 
\exp(i \sum_m {\bf c}_m \cdot M_1({\vec J}_m)) \Psi_{\vec J},
\ee
where again the dot product is defined with respect to an appropriate metric
on ${\cal L}_{\cal C}$.

Physical states correspond to proper junctions ${\vec J}_{\cal P}$
with $\sum_m p_m = \sum_m q_m = 0$. 
The global structure of the gauge group is determined
by finding the elements in ${\cal C}$ which
act trivially on all such junctions. In other words, we seek $\bf c$ 
such that
\be
\exp(i {\bf c} \cdot M_1({\vec J}_{\cal P})) = 1.
\label{centnull}
\ee  
$\bf c$ may be represented by a junction ${\vec J}_c$ of the form
\be
{\vec J}_c = \sum_m {\vec J}_{c,m} = \sum_m P_m{\vec w}_{p,m} + Q_m{\vec w}_{q,m},
\label{form}
\ee
for which
\be
{\bf c} = M_2({\vec J}_c)
\ee
Thus we seek ${\vec J}_c$ such that
\be
\exp(i M_2({\vec J}_c) \cdot M_1({\vec J}_{\cal P})) = 1
\label{expr}
\ee
for all ${\vec J}_{\cal P}$.

A natural guess for a solution of (\ref{expr}) is that
${\vec J}_c$ is a null junction (defined in section 2).
The dot product of a null junction with any other junction is zero when
computed with respect to the metric on the junction lattice (see \cite{DZ}). 
The same is not manifestly true after mapping with $M_2$ and
computing the dot product with respect to the metric on ${\cal L}_C$.
However we will argue below that this guess is correct  
for an appropriate choice of metric on ${\cal L}_C$.  
It is not be true for all choices of metric, since
the element of ${\cal C}$ represented by a junction ${\vec J}_c$ 
depends on this choice,  and is otherwise determined only up 
to an automorphism of $\cal C$. The proper null junctions,  
since they are proper,  are mapped by $M_2$ to the identity element of ${\cal C}$.
To get non-trivial elements of ${\cal C}$ we shall have to consider improper null
junctions which are fractions of the proper null junctions. 
The subgroup ${\cal H}$ of ${\cal C}$
which acts trivially on all physical states is then given by 
\be
{\cal H} = {\cal N}/{\cal N}_{\cal P},
\label{nnp}
\ee
where ${\cal N}_{\cal P}$ is the lattice of proper null junctions and 
${\cal N}$ is a lattice of null junctions generated by the fractions of
proper null junctions with integer $P_m$ and $Q_m$.

To show (\ref{nnp}), it is more convenient to work
with the maps ${\tilde M}$ (as in (\ref{tilmaps}) ).
Since ${\vec J}_{\cal P} = \sum_m(p_m\vw_{p,m} + q_m\vw_{q,m} \cdots)$ 
is a proper junction,  
(\ref{expr}) may be re-written as 
\be 
\exp\left( i \sum_m \tM ({\vec J}_{c,m})  
\cdot {\tilde M}_2(p_m{\vec w}_{p,m} + q_m{\vec w}_{q,m})\right) = 1
\label{propact}
\ee
(\ref{propact}) must hold for all $p_m$ and $q_m$ satisfying 
$\sum_m p_m = \sum_m q_m = 0$.
The solutions of (\ref{propact}) are junctions ${\vec J}_c$ with the property that 
\be
\exp \left( i\tM ({\vec J}_{c,m}) \cdot {\tilde M}_2({\vec w}_{p,m}) \right) = z_p
\label{propone}
\ee
and 
\be
\exp \left( i \tM({\vec J}_{c,m})\cdot {\tilde M}_2({\vec w}_{q,m}) \right) = z_q
\label{proptwo}
\ee
for all $m$ for which $\vw_{p,m}$ or $\vw_{q,m}$ are defined\footnote{
$\vw_{p,m}$ or $\vw_{q,m}$ are not both defined for $I_n$ fibers, corresponding
to collections of 7-branes with the same charges.} 
There is no sum on the index $m$ in the above expressions and $z_p$ and $z_q$ 
are independent of the index $m$. 
Due to the independence of (\ref{propone})
and (\ref{proptwo}) on $m$,   
the center of the universal
cover, ${\cal C}=\prod_m Z_{(N_m)} \times Z_{(M_m)}$,  must have a trivially
acting subgroup ${\cal H}$ of the form $Z_l \times Z_r$.  

Given a choice of metric,
it very easy to construct a lattice of junctions 
representing elements of the trivially acting subgroup ${\cal H}$ 
of ${\cal C}$. For instance, one can first construct the lattice $\Gamma_{P}$ 
of junctions ${\vec J}_c =  \sum_m (P_m {\vec w}_{p,m} + 
Q_m {\vec w}_{q,m})$ for which  the quantities
\be
{\tilde M}_2 (P_m{\vec w}_{p,m} + 
Q_m{\vec w}_{q,m}) \cdot {\tilde M}_2({\vec w}_{p,m}) 
\label{lpropone}
\ee
and
\be
{\tilde M}_2(P_m{\vec w}_{p,m} + 
Q_m{\vec w}_{q,m}) \cdot {\tilde M}_2({\vec w}_{q,m}) 
\label{lproptwo}
\ee
are independent of the index $m$, and equal to an integer multiple of $2\pi$.
This clearly satisfies (\ref{propone}) and (\ref{proptwo}).
Junctions satisfying this condition are proper and represent the identity 
element of ${\cal C}$. Let us call the lattice of such junctions
$\Gamma_{\cal P}$.  
By considering fractions of these 
junctions such that $P_m$ and $Q_m$ remain integer, one preserves
the independence of (\ref{lpropone}) and (\ref{lproptwo}) on
the index $m$,  but gets representatives of all elements of $\cal C$   
which act trivially on physical states.  
These improper junctions generate a lattice $\Gamma$ containing 
$\Gamma_{\cal P}$.  In terms of these two lattices,
\be
\pi^1({\tilde G}) = {\cal H} =\frac{\Gamma}{\Gamma_{\cal P}}
\ee

In some cases the lattice of null junctions ${\cal N}$ is of the type 
$\Gamma$ described above, but not always.  
Nevertheless there exists 
a choice of metric such that the trivially acting components of ${\cal C}$ are
represented by ${\vec J}_c$ in ${\cal N}$.  
In general,  for the appropriate choice of metric, 
the elements ${\vec J}_c$ in ${\cal N}$
satisfy the weaker conditions (\ref{propone}) and (\ref{proptwo}).

The lattice of null junctions has a proper sublattice 
${\cal N}_{\cal P}$ corresponding to the identity element of ${\cal C}$.
Generators of the lattice of proper null junctions
can be obtained from string loops encircling
all the 7-branes,  with charges $(1,0)$ or $(0,1)$ in
the upper half-plane. 
Pulling the lower
half of the loop through the 7-branes (as in figures 4,8),  so that the entire
junction lives in the upper half plane,  gives a junction
of the form $\sum_m P_m{\vec w}_m + Q_m{\vec w}_m$.
Since these junctions map to the identity element of 
$\cal C$ under ${\tilde M}_2$,
(\ref{propone}) and (\ref{proptwo})
are trivially satisfied,  with $z_p = z_q =1$.

The full lattice of null junctions ${\cal N}$ is obtained
by taking fractions of the junctions in ${\cal N}_{\cal P}$
such that $P_m$ and $Q_m$ remain integer.
The resulting lattice has has dimension $2$, as desired to get 
$\pi^1(\tilde G)$ of the expected form $Z_l \times Z_r$.
In appendix A. we show  that
the fractional null junctions satisfy (\ref{propone}) and (\ref{proptwo})
for the appropriate choice of metric. 

Note that the lattice of proper null junctions ${\cal N}_{\cal P}$ 
does not depend 
on the type of Kodaira singularities in the elliptic surface,  which can 
be changed by moving 7-branes around within the string loop which defined
the proper null junction.  However the lattice ${\cal N}$ which includes 
the fractional null junctions does depend on the type of Kodaira singularities,  
since the allowed  fractions with integer $P_m$ and $Q_m$ depend on the 
7-branes which are within each collection labelled by $m$. 

\noindent In summary:

\noindent{\it $\pi^1({\tilde G})$ is 
the quotient of the lattice of null junctions ${\cal N}$ by the lattice of 
proper null junctions ${\cal N}_{\cal P}$.} 

\be
\pi^1({\tilde G}) = {\cal H} = \frac{ {\cal N}  }{ {\cal N}_{\cal P} } 
\ee
We wish to emphasize that the null junctions determine not only $\pi^1({\tilde G})$,  but also the
entire global structure of ${\tilde G}$.  To obtain the full global structure of 
${\tilde G}$,
one requires the elements of the center of the 
universal cover of ${\tilde G}$ which act trivially
in all representations of ${\tilde G}$.  These elements are the image of the null junctions
${\cal N}$ under the map $M_2$.

\subsection{Examples for rational elliptic surfaces}

To clarify some of the above statements,  we illustrate 
them for some particular examples.
Consider a rational elliptic surface with Kodaira fibers $III$, $III$, and $I_0^*$.
The universal cover of ${\tilde G}$ is $SU(2) \times SU(2)\times Spin(8)$, with center 
${\cal C} = Z_2 \times Z_2 \times (Z_2\times Z_2)$.  
Arranging these fibers from left to right on the real axis,  as in 
figure 5, we shall take $m=1,2$ to correspond to the two $III$ fibers and 
$m=3$ to correspond to the $I_0^*$ fiber.  In terms of its decomposition to 
$I_1$'s (individual 7-branes)  this configuration is $(AAC)(AAC)(AAAABC)$ - see table 1(a).
 
Let us first consider $m=1$. Then, from table 1(b), 
\bea
{\vec w}_{p,1} = (\frac{1}{2},\frac{1}{2},0) \nonumber \\
{\vec w}_{q,1} = (\frac{1}{2},\frac{1}{2},-1).
\eea
The lattice spanned by these junctions becomes $Z_2$ (the center of $SU(2)$) 
upon quotienting by the proper junctions.
Furthermore
\be
M_2({\vec w}_{p,1}) = M_2({\vec w}_{q,1}) = 1 \  mod  \ 2
\ee
The junction $P_1{\vec w}_{p,1} + Q_1 {\vec w}_{q,1}$
represents an element of the center determined by the map $M_2$.  This
element acts as follows in 
a representation containing a proper junction with asymptotic charges $p_1,q_1$; 
\bea
\Psi_{{\vec J}_{\cal P}}\rightarrow & \exp \left( i\tM(P_1\vw_{p,1} +Q_q\vw_{q,1}) \cdot \tM(p_1\vw_{p,1} + q_1\vw_{q,1}) \right) \Psi_{{\vec J}_{\cal P}} = \nonumber \\ 
& (-1)^{(P_1+Q_1)(p_1+q_1)} \Psi_{{\vec J}_{\cal P}}
\eea

\begin{figure}
\begin{center}
\epsfig{file=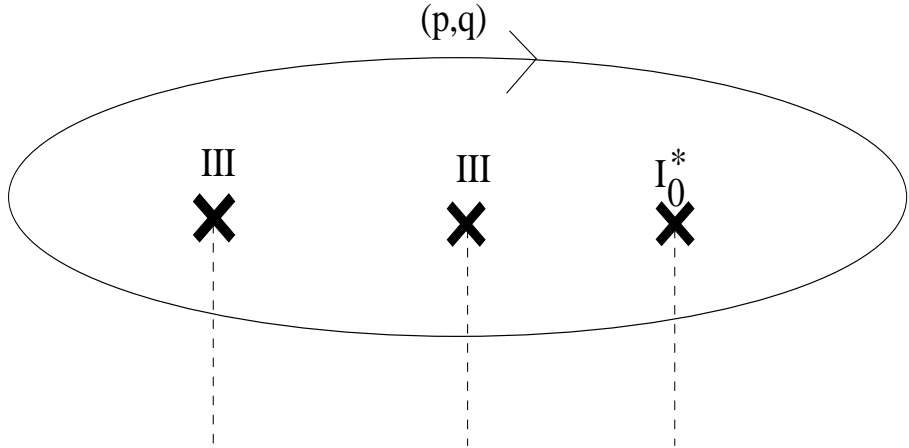,height=6cm,width=12cm}
\end{center}
\caption{ proper null junctions of a rational elliptic surface.     }
\end{figure}

For $m = 3$  one has
\bea
{\vec w}_{p,3} = (0,0,0,0,\frac{1}{2},\frac{1}{2}) \nonumber \\
{\vec w}_{q,3} = (\frac{1}{2},\frac{1}{2},\frac{1}{2},\frac{1}{2}, 
-\frac{3}{2}, -\frac{1}{2}).
\eea
The lattice spanned by these junctions becomes $Z_2 \times Z_2$
(the center of $Spin(8)$)  upon quotienting by the proper junctions.
We shall represent an element of the center by a pair of integers
$(j,k)$, each defined modulo 2.
Then
\bea
M_2({\vec w}_p) = (1,0)\nonumber \\
M_2({\vec w}_q) = (0,1)
\eea
The junction $P_3{\vec w}_{p,3} + Q_3 {\vec w}_{q,3}$
represents an element of the center determined by the map $M_2$.  In 
a representation containing a proper junction ${\vec J}_{\cal P}$ with asymptotic 
charges $p_3$ and $q_3$, this
element of the center acts as follows;  

\be
\Psi_{{\vec J}_{\cal P}} \rightarrow (-1)^{P_3p_3} (-1)^{Q_3q_3} 
\Psi_{{\vec J}_{\cal P}}
\ee 

The lattice of proper null junctions is generated by
\be
{\vec N}_1 = -{\vec w}_{p,1} - {\vec w}_{q,1} - {\vec w}_{p,2} + {\vec w}_{q,2} + 2 {\vec w}_{p,3}
\label{nulljctone}
\ee
and
\be
{\vec N}_2 = {\vec w}_{p,1} - {\vec w}_{q,1} - {\vec w}_{p,2} - {\vec w}_{q,2} + 2 {\vec w}_{q,3}
\label{nulljcttwo}
\ee
corresponding to the string loops in figure 5 with $(p,q) = (1,0)$ and $(0,1)$ 
respectively.  (\ref{nulljctone}) and (\ref{nulljcttwo}) are obtained 
upon pulling the lower half of the loop through the 7-branes and into the upper 
half plane (see figures 4,8).
Modulo proper null junctions,  there is precisely one fractional 
null junction with integer $P_i$ and $Q_i$. It is given by 
\be
\frac{1}{2} ({\vec N}_1 + {\vec N}_2) = 
-{\vec w}_{q,1}
 - {\vec w}_{p,2} + {\vec w}_{p,3} + {\vec w}_{q,3}.
\ee
This corresponds 
to the element $-1 \times  -1 \times (-1 \times -1)$  
in ${\cal C} = Z_2 \times Z_2 \times (Z_2 \times Z_2)$ 
In a representation containing a {\it proper} junction
of the form 
\be
{\vec J}_{\cal P} = \sum_{i=1}^3 (p_i{\vec w}_{p,i} + q_i{\vec w}_{q,i} + a_{k,i}{\vec w}_{k,i}),
\ee
this element of the center acts as follows;
\be
\Psi_{{\vec J}_{\cal P}} \rightarrow 
(-1)^{(q_1 + p_1)}(-1)^{(p_2 + q_2)}(-1)^{p_3}(-1)^{q_3} \Psi_{{\vec J}_{\cal P}}
\label{ztwoaction}
\ee
This action is trivial when $\sum_m p_m = \sum_m q_m =0$.
Thus there is a $Z_2$ subgroup of the center of the universal cover which acts 
trivially when $\sum_i p_i = \sum_i q_i =0$.  
The semi-simple part of the gauge group is then
\be
{\tilde G} = \frac{SU(2) \times SU(2)\times Spin(8)}{Z_2},  
\ee
where the $Z_2$ action is given by (\ref{ztwoaction}).

Let us briefly consider another example.   Consider the rational elliptic surface 
with an $I_3, I_1,$ 
and $E_6$ singularity.  The universal cover is $SU(3) \times E_6$ with 
center $Z_3 \times Z_3$.  One can arrange these singularities from left to right
on the real axis in the base,  with indices $m=1,2$ and $3$ labelling
the $I_3, I_1$ and $E_6$ fibers respectively. The monodromies are (up to an overall
$SL(2,Z)$ conjugation) 
\bea
\nonumber \\
M_1 = \pmatrix{1 & -3 \cr 0 &1\cr}, M_2 = \pmatrix{1 & -0 \cr 1 &1\cr},
M_3 = \pmatrix{-2& 3 \cr -1 & 1\cr}. \\
\nonumber 
\eea
Note that this implies $q_1 = p_2 =0$.  
It is easy to show (see the appendix) that $\vw_{q,2}$ and $\vw_{q,3}$
are proper, while $3\vw_{p,1}$ and $3\vw_{p,3}$ are
minimal proper junctions,  meaning that they can not be divided by any integer
$>1$ and remain proper. 
This accounts for the fact that the center of the universal cover is
$Z_3 \times Z_3$.   
We choose a metric such that
\bea
\tM (\vw_{p,1}) \cdot \tM (\vw_{p,1}) = \frac{2\pi}{3} \nonumber \\
\tM (\vw_{p.3}) \cdot \tM (\vw_{p,3}) = \frac{2\pi}{3}
\label{dots}
\eea
Since the junctions $\vw_{q,2}$ and $\vw_{q,3}$ are proper, all dot products involving
them are are integer multiples of $2\pi$.
The proper null junctions are
\be
\vec N_{(\tilde p,\tilde q)} = 3 \tilde q \vec w_{p,1} + 
(3 \tilde q - \tilde p) \vec w_{q,2} -3 \tilde q \vec w_{p,3} + 
(\tilde p - 3 \tilde q) \vec w_{q,3}
\ee
where $(\tilde p, \tilde q)$ indicates the charge in the upper half plane of
the corresponding string loop surrounding all three singular fibers. 
The quotient lattice ${\cal N}/{\cal N}_{\cal P}$ 
is $Z_3$ and is generated by the fractional null junction 
\be
\vec N = \frac{{\vec N}_{(0,1)}}{3} = \vec w_{p,1} + \vec w_{q,2} - \vec w_{p,3} - \vec w_{q,3}
\label{zthregen}
\ee
In a representation containing the proper junction
$\vec J_{\cal P} = \sum_m p_m \vw_{p,m} + q_m \vw_{q,m} + \cdots$,
the element of the center associated with 
(\ref{zthregen}) acts as follows;
\bea
\Psi_{\vec J_{\cal P}} \rightarrow & 
\exp\left(i\tM({\vec N}) \cdot \tM(\sum_m p_m \vw_{p,m} + q_m \vw_{q,m})\right) 
\Psi_{\vec J_{\cal P}} = \nonumber \\
& \exp \left( \frac{2\pi i}{3} (p_1 + p_3) \right) \Psi_{\vec J_{\cal P}}.
\label{zthract}
\eea
Since $p_2=0$,  we must have $p_1 + p_3 = 0$,  and this action is 
trivial.  Thus we find
\be
\tilde G = \frac {SU(3) \times E_6}{Z_3}
\ee
where the $Z_3$ action is given by (\ref{zthract}).

\subsection{Gauge group for an elliptic K3.}

In \cite{AM} the gauge group for F-theory compactifications on 
a $K3$ was computed at the boundary of moduli space at which a $K3$ 
degenerates to two rational elliptic surfaces intersecting over an elliptic
curve.  The   
spectrum in this case is 
consistent with a gauge group which is a direct product 
of groups computed for each rational elliptic surface.
The means of computing $\pi^1({\tilde G})$ in \cite{AM}
differs from ours,
as it involves identifying $\pi^1({\tilde G})$ with the torsion component of
the Mordell-Weil lattice $T(\Phi)$.  Since we have identified  $\pi^1(\tilde G)$ with
${\cal N}/{\cal N}_{\cal P}$,  our results  explains the equivalence of 
${\cal N}/{\cal N}_{\cal P}$ with $T(\Phi)$ which was observed for rational elliptic 
surfaces in \cite{FY}.  Our methods are not restricted to rational elliptic surfaces however.
They are equally applicable to generic elliptic K3's and, as we shall 
later argue, to some elliptically fibered three-folds. 

The computation of the gauge group in \cite{AM} considers K3's 
which have degenerated to a pair of  rational
elliptic surfaces ${\cal R}_i$, and does  not take into account certain multiplets 
which become infinitely massive in this limit. 
In this limit,  the $P^1$ base of the $K3$ pinches to
two $P^1$'s meeting at a point.  The spectrum considered in
\cite{AM} consists of elements of $H_2(X,Z)$ within each rational elliptic surface.
These elements of $H_2(X,Z)$ project to strings within each $P^1$ but not stretching 
between them.  
Strings stretching between the $P^1$ bases correspond to elements of $H_2(K3,Z)$ 
but not of $H_2({\cal R}_i,Z)$.  These strings are necessary to account
for the difference between the second Betti number of a K3, $b_2 = 22$,  and twice that
of a rational elliptic surface,  which has $b_2 = 10$.
In the dual heterotic description on $T^2$, they correspond to 
strings winding cycles of the $T^2$ whose radii are becoming infinite in the
degenerate limit.  (see \cite{Imamura}
for a discussion of string junctions and heterotic-IIB duality).
The algebra associated with strings junctions on a rational
elliptic surface is ${\hat E}_9$ (see \cite{DHIZ2}),  whereas the Narain lattice of
the heterotic theory on a torus is equivalent to the root lattice of $E_{10} \oplus
E_{10}$.  If one deforms the K3 away from the stable degeneration to a pair of rational 
elliptic surfaces, one can not 
neglect strings stretching between the two $P^1$ bases,  as they are no longer 
infinitely massive.

\begin{figure}
\begin{center}
\epsfig{file=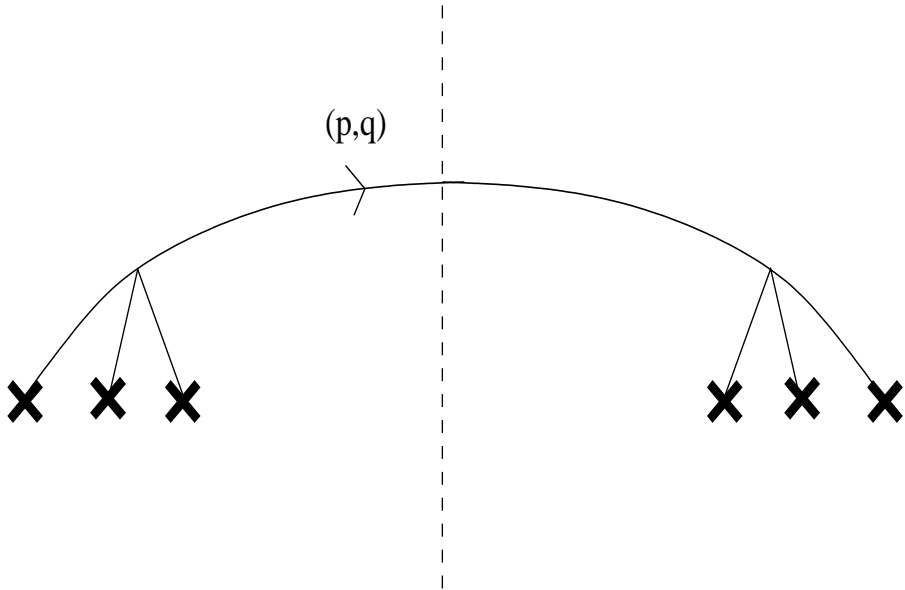,height=8cm,width=12cm}
\end{center}
\caption{A string junction stretched between two sets of 12 7-branes 
in the base of an elliptic K3.  In the limit in which the K3 degenerates 
to two rational elliptic surfaces,
the dashed line (an $S^1$ equator on $P^1$) separating the two sets of 
7-branes shrinks to a point.     }
\end{figure}

\begin{figure}
\begin{center}
\epsfig{file=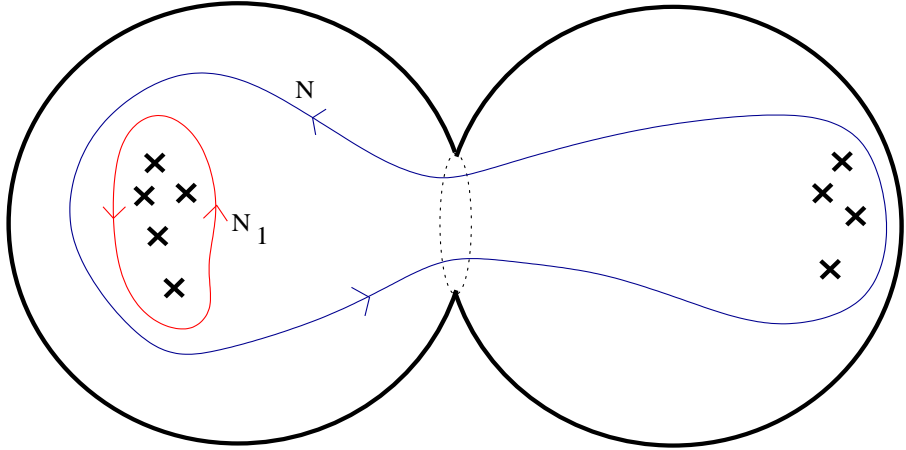,height=6cm,width=12cm}
\end{center}
\caption{ Pinching of the base of an elliptic K3 near the stable degeneration 
to a pair of intersecting rational elliptic surfaces.  $N_1$ is equivalent
to the $0$ junction on the base of a rational elliptic surface,  but not
on the base of K3.  $N$ is equivalent to the 0 junction on the base of
the K3,  and its fractions determine $\pi^1(\tilde G)$.} 
\end{figure}

Near the degeneration to an intersecting pair of rational elliptic 
surfaces ${\cal R}_i$, the 7-branes on the $P^1$ base of a K3 
may be split into two groups of 12 with unit total monodromy.  These groups of 7-branes
live in the base of each ${\cal R}_i$ in the degenerate limit.
The proper null junctions of a generic K3 are string loops surrounding  {\it all}
$24$ 7-branes.  $\pi^1({\tilde G})$ is generated by fractions of
these null junctions having integer $P_m$ and $Q_m$.
The gauge group has the form $(G_1 \times G_2) / \Gamma$
with $G_1$ and $G_2$ simply connected and $\Gamma \equiv {\cal N}/{\cal N}_{\cal P}$, 
but is generically not a direct product of the form $G_1/\Gamma_1 \times G_2/\Gamma_2$ 
as found in \cite{AM}. The $\Gamma_i$ computed in \cite{AM} are isomorphic to 
${\cal N}_i/{\cal N}_{{\cal P},i}$,  where ${\cal N}_i$ are null junctions in
the base of ${\cal R}_i$. These junctions are related to
string loops surrounding the relevant set of $12$ 7-branes.
The elements in the center of the universal cover represented by junctions in
${\cal N}_i$ do not act trivially on junctions stretching between the 
two groups of 12 7-branes,  since such junctions have a non-zero asymptotic $(p,q)$
charges within each of the $P^1$ components of the base (see figure 6).
Recall that the condition for null junctions to represent trivially acting elements
in the center of the universal cover requires that the asymptotic charges of
all states be zero.  Note also that on a K3,  string loops encircling just $12$ 
of the branes (with unit total monodromy) are not really null (see figure 7),
since they are inequivalent to the $0$ string junction and represent
a a nontrivial element of $H_2(K3,Z)$.   In conclusion, for a $K3$  

\noindent {\it 
$\pi^1(G) = {\cal N}/{\cal N}_{\cal P}$ where ${\cal N}_{\cal P}$ are null 
junctions arising from string loops surrounding all $24$ 7-branes. }

\noindent In the next section we compute the global structure of the gauge group for a K3,
including all the $U(1)$ factors as well.

\section{The global structure including $U(1)$ factors.}

Thus far we have only considered the global structure of the semisimple part of 
the gauge group ${\tilde G}$.  One can readily extend the discussion to include
the $U(1)$ factors.  Consider a general junction characterized by Dynkin labels
$a_{i,m}$ and charges $p_m$ and $q_m$ with $m= 1 \cdots r$.  The $U(1)$ charges
are related to the $p_m$ and $q_m$ charges.
Since we require $\sum_m p_m = \sum_m q_m = 0.$,  there are $2r-2$ independent 
$p_m$ and $q_m$ charges.  Of these,  only $2r - 4$ are conserved.  This is because
there is dimension $2$ lattice of null junctions,  which are topologically equivalent
to the zero junction.  For a K3, the rank of the entire gauge group, including
both ${\tilde G}$ and the $U(1)$ factors,  is $20$,  which is what
one expects from duality with the heterotic string on $T^2$.  

One can write the $2r-4$ conserved $U(1)$ charges ${\cal Q}_s$ as linear combinations of 
the independent $p_m$ and $q_m$ which are invariant under
the additions of null junctions. 
The $U(1)$ factors in the gauge group act as follows in representations containing
junctions ${\vec J}$  with charges ${\cal Q}_s$;
\be
\Psi_{\vec J} \rightarrow  \exp(i\theta_sQ_s) \Psi_{\vec J}.
\ee  
For certain values of
$\theta_s$,  this may be the same as the action of some element in the center
of ${\tilde G}$. The results of the previous section give the action of the
center of ${\tilde G}$,  so it is easy to compute the global structure
when one includes all the $U(1)$ factors.  

As a simple example,  consider a $K3$ with four $I_0^*$ fibers.
Locally,  the gauge group is $U(1)^4 \times Spin(8)^4$.
The lattice of proper null junctions is generated by
\be
{\vec N}_1 = 2{\vec w}_{p,1} - 2{\vec w}_{p,2} + 2{\vec w}_{p,3} - 2{\vec w}_{p,4}  
\ee
and
\be
{\vec N}_2 = 2{\vec w}_{q,1} - 2{\vec w}_{q,2} + 2{\vec w}_{q,3} - 2{\vec w}_{q,4}  
\ee
Therefore the conserved $U(1)$ charges are 
\bea
Q_1 = p_1 + p_2 = -p_3 - p_4 \nonumber \\
Q_2 = p_1 + p_4 = -p_2 - p_3 \nonumber \\
Q_3 = q_1 + q_2 = -q_3 - q_4 \nonumber \\
Q_4 = q_1 + q_4 = -q_2 - q_3
\eea

The center of $Spin(8)^4$ is $(Z_2 \times Z_2)^4$.
Using the results of the previous section, the action of the center 
in a representation containing the proper junction 
${\vec J}_{\cal P} = \sum_m \left(  p_m {\vec w}_{p,m} 
+ q_m {\vec w}_{q,m} + a_{i,m}{\vec w}_{i,m}  \right)$ 
is;
\be
\Psi_{{\vec J}_{\cal P}} \rightarrow \exp \left( i\pi \sum_m (l_m p_m + r_m q_m)
\right)
\Psi_{{\vec J}_{\cal P}}
\ee
where for each $m$, $l_m$ and $r_m$ are integers indicating an element of 
$Z_2 \times Z_2$.
There is a $Z_2 \times Z_2$ subgroup of $(Z_2 \times Z_2)^4$ 
which acts trivially
on all ${\vec J}_{\cal P}$ with $\sum_m p_m = \sum_m q_m = 0$.
This subgroup is generated by $(r_m, l_m) = (1,0)$ for all $m$ and
$(r_m,l_m) = (0,1)$ for all $m$.  
The semisimple part of the gauge group is therefore
\be
{\tilde G} = \frac{Spin(8)^4}{Z_2 \times Z_2}
\ee

Among the {\it non}-trivially acting elements of $(Z_2 \times Z_2)^4$
are some which are equivalent to elements in the $U(1)$ parts of
the gauge group;
\be
\exp(i \theta_i Q_i) =   exp( i\pi \sum_m (l_m p_m + r_m q_m))
\ee
A $U(1)$ factor with $\theta_i = \pi$ for any $i$ is equivalent
to a non-trivially acting  element in the center of $Spin(8)^4$.
Thus we find that the full gauge group is
\be
G = \frac{U(1)^4 \times Spin(8)^4}{Z_2^6}
\ee

\section{Elliptically fibered Calabi-Yau three-folds and $\pi^1(G)$.} 

String junction methods may also be applicable to the determination of 
the global structure of the gauge group for F-theory compactifications
on elliptic Calabi-Yau three-folds.  
Some new complexities arise
in this case due to singularities of the discriminant curve over which
different Kodaira types collide. 
The relation of $\pi^1({\tilde G})$ to the torsion component of the Mordell-Weil 
lattice of an elliptic Calabi-Yau three-fold was shown to hold in certain 
special instances in \cite{AM}. However the Mordell-Weil lattices of elliptic 
Calabi-Yau three-folds have not been generally classified,
and the relation of its torsion component to $\pi^1({\tilde G})$ is un-proven.
Our purpose in this section is to suggest a general approach to obtain 
the global structure of the gauge group using string junctions.

Consider a K3 fibered Calabi-Yau three-fold, of ``perturbative'' type, for
which the K3 fibers become more singular at isolated points  $P$ in the $P^1$ base due
to the collision of Kodaira fibers.  In this case 
we propose to determine the global structure of the gauge group
as follows.  One first computes the group $G$ 
associated to a generic K3 fiber using the methods of the previous sections.
The actual gauge symmetry {\it algebra} ${\cal G}$ may be a sub-algebra of that of
$G$,  and in some instances is non-simply laced \cite{AG, BIKMSV, AKM}.  
We conjecture that the gauge symmetry group is simply a subgroup of $G$ with 
the algebra ${\cal G}$.  This will be true provided that the string junction
lattice of the Calabi-Yau is that of a generic $K3$, even if the gauge symmetry 
algebra is a sub-algebra of that of a generic $K3$. 
This requires
that junctions  corresponding to vector multiplets in  compactifications 
on a K3 $X$ do not disappear from the spectrum in
compactification on a Calabi-Yau $Y$ for which $X$ is the generic fiber,  
but may instead correspond to hypermultiplets.  

There is good evidence that this is the case.  For instance, massless 
vector multiplets arise from membranes wrapping vanishing cycles in $H_2(Y,Z)$,  
whereas massless hypermultiplets arise from membranes wrapping unions of rational 
curves \cite{AKM} in the blowup of the elliptic fiber over the singular points
where components of the discriminant curve collide.  These unions of rational
curves do not necessarily correspond to elements of $H_2(Y,Z)$.  
A discussion of hypermultiplets which are not contained in 
$H_2(Y,Z)$ appeared originally in \cite{KV}.
In the non-simply laced cases,  there is a monodromy which acts on the Dynkin
diagram associated to $X$ \cite{AG, BIKMSV, AKM}.  This monodromy removes some
roots from the gauge symmetry algebra,  however there are hypermultiplets
arising from unions of rational curves in the elliptic fiber which are exchanged 
under the monodromy.

The string junction lattice for a Calabi-Yau three-fold is currently being investigated
in (see \cite{US}).  
String junctions on an elliptically fibered three-fold differs somewhat 
from those  on an elliptic surface
(a two-fold), since the string junctions live in a base of four 
real dimensions rather than two.
The discriminant locus
is a collection of complex curves,  rather than points.  Strings 
may end on these curves,  and in addition to the equivalence 
relations between strings related by smooth deformations and
Hanany-Witten transitions,  there is an additional equivalence
between strings related by sliding endpoints along the discriminant 
curve. A subtlety arises due to the singular points of the discriminant 
curve,  where different smooth components meet or intersect.  
A well defined string junction lattice can be  obtained
by finding the equivalence classes of string junctions in the space 
$B - {s_{\alpha}}$  where $B$ is the base of
the elliptically fibered Calabi-Yau 
and $s_{\alpha}$ are the singular 
points of the discriminant curve at which Kodaira types collide.

Consider again a  K3-fibered Calabi-Yau  $X$,  for which the K3
is fibered over a lower $P^1$  and the $K3$ fiber is itself
an elliptic fibration over an upper $P^1$. 
We assume the $K3$ fiber 
degenerates at isolated points ${\cal Q}_s$ in
the lower $P^1$, over which smooth components of the discriminant
curve collide. 
Via an equivalence relation,  any string junction can be brought
entirely within an upper $P^1$, over a point ${\cal Q}$ in the lower
$P^1$.  This suggests that the string junction lattice is that 
of a generic $K3$ fiber, up to a possible quotient by the
action of a monodromy as ${\cal Q}$ encircles a point ${\cal Q}_s$ 
at which the  $K3$ becomes more singular due to the collision 
of Kodaira fibers.  Within a generic upper $P^1$,  string junctions 
end at points 
${\cal Y}_i$ where the discriminant curve intersects the upper $P^1$.
As $Q$ encircles $Q_s$,  the ${\cal Y}_i$ corresponding to 
Kodaira fibers which collide at ${\cal Q} ={\cal Q}_s$ move around each other,
sweeping out a braid representation of a knot or link. 
This generates a natural action of the braid group on string junctions;
\be
{\vec J}\rightarrow {\hat B}{\vec J}
\ee
where ${\hat B}$ is an element of the braid group, and ${\vec J}$
is a string junction defined over the point ${\cal Q}$.
The junctions ${\vec J}$ and ${\hat B}{\vec J}$ are
equivalent in the three-fold,  even if they are 
inequivalent when restricted to a point ${\cal Q}$.
If $\hat B$ acts non-trivially,  then any
junction is equivalent to one with endpoints on
a subset of the points ${\cal Y}_i$ 
One expects $\hat B$  to act non-trivially if the 
braid is non-minimal,  i.e. if there is another
braid representation of the same knot with a fewer number of 
strands.   

Given some of the above statements about the hypermultiplet spectrum for 
the Calabi-Yau compactifications,  one would expect that the 
the braid which arises when ${\cal Q}$ encircles 
${\cal Q}_s$ is always minimal so that the 
quotient acts trivially, although we will not be able to prove this here.  
Related questions concerning the properties of algebraic knots
are discussed in \cite{US}.

%%%%%%%%%%%%%%%%%%%%%%%%%%%%%%%%%%%%%%%%%%%%%%%%%%%%%%

\section*{Acknowledgements}

I am grateful to O. Ganor, A. Grassi, B. Ovrut,
T. Pantev and B. Zwiebach for useful discussions.
This research was supported in part by funds provided
by the U.S. Department of Energy under cooperative 
research agreement DF-FC02-94ER40818.

%%%%%%%%%%%%%%%%%%%%%%%%%%%%%%%%%%%%%%%%%%%%%%%%%%%
\section{Appendix.}

\begin{figure}
\begin{center}
\epsfig{file=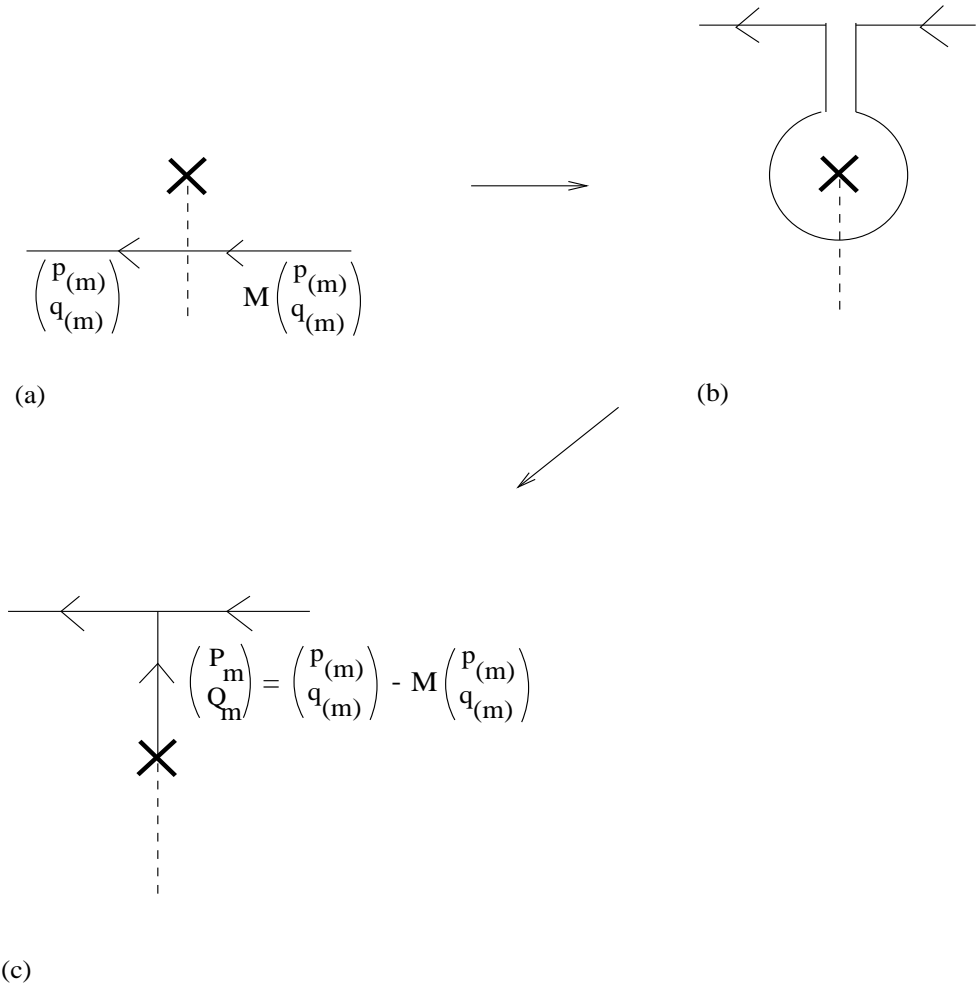,height=13cm,width=13cm}
\end{center}
\caption{ Hanany-Witten transition upon pulling a segment of 
a string loop through a collection of 7-branes. }
\end{figure} 
 
We will show below that the null junctions 
with integer $P_m$ and $Q_m$ 
obtained from fractions of proper null junctions 
satisfy (\ref{propone}) and (\ref{proptwo}).

Consider a proper null junction obtained by starting with
a string loop surrounding
all 7-branes and deforming it into the upper half plane by
pulling the lower part of the loop through all the 7-branes.
Suppose also that the charge of the string loop is $(\tilde p,\tilde q)$
in the upper half plane.
The resulting junction ${\vec N}_{(\tilde p,\tilde q)}$ has the form
\be
{\vec N}_{(\tilde p,\tilde q)} = \sum_m {\vec J}_{c,m} = 
P_m {\vec w}_{p,m} + Q_m {\vec w}_{q,m}
\ee
The full (one dimensional) lattice of null junctions proportional to 
${\vec N}_{(\tilde p,\tilde q)}$ 
is generated by
\be
{\vec N}_{(\tilde p,\tilde q)} / L_{\tilde p,\tilde q} 
\ee
where
\be
L_{\tilde p,\tilde q} = gcd(P_1,P_2,\cdots, Q_1,Q_2,\cdots)
\ee
and $gcd$ indicates the greatest common divisor. 
We wish to show that (\ref{propone}) and (\ref{proptwo}) are satisfied by
the null junctions in this lattice.  In other words,  we wish to show that
\be
\exp \left( i \frac{1}{ L_{\tilde p,\tilde q}   } 
\tilde M_2(P_m {\vec w}_{p,m} + Q_m {\vec w}_{q,m} )
\cdot {\tilde M}_2({\vec w}_{p,m}) \right) = z_p
\label{agpropone}
\ee
and 
\be
\exp \left( i \frac{1}{ L_{\tilde p,\tilde q}   } \tilde M_2 
( {\vec w}_{p,m} + Q_m {\vec w}_{q,m} )
\cdot {\tilde M}_2({\vec w}_{q,m}) \right) = z_q
\label{agproptwo}
\ee
where there is no sum on the index $m$ in the above expressions and $z_p$ and $z_q$ 
are independent of the index $m$. 
It will suffice to show (\ref{agpropone}) and (\ref{agproptwo}) for 
a generating pair of ${\vec N}_{(\tilde p, \tilde q)}$  
with $gcd(\tilde p, \tilde q) = 1$,  such as $(1,0)$ and $(1,0)$,.

Since the junction ${\vec J}_{c,m} = P_m  {\vec w}_{p,m} + Q_m {\vec w}_{q,m}$ 
(with no sum on $m$) is proper, 
\be
\tilde M_2(P_m {\vec w}_{p,m} + Q_m {\vec w}_{q,m} ) 
\cdot \tilde M_2 ( {\vec w}_{p,m} )  = 2\pi j_m
\ee
and 
\be
\tilde M_2(P_m {\vec w}_{p,m} + Q_m {\vec w}_{q,m} ) 
\cdot {\tilde M}_2({\vec w}_{q,m})  = 2\pi k_m
\ee
for integer $j_m$ and $k_m$.  
In the absence of constraints on these integers (\ref{agpropone}) and 
(\ref{agproptwo}) would not be satisfied.  Fortunately,  there are 
constraints.
To see this let us consider the effect of pulling a segment of a string loop below
the $m$'th collection of 7-branes into the upper half plane.
The charges of this string segment are  $(p_{(m-1)},q_{(m-1)})$ to the left of 
the branch cut and $(p_{(m)}, q_{(m)})$ to the right of the branch cut, 
(see figure 8), where these are related by the $SL(2,Z)$ monodromy $M_m$,
\be
\pmatrix{p_{(m+1)} \cr q_{(m+1)} \cr} = M_m \pmatrix{ p_{(m)}\cr q_{(m)} \cr} 
\ee
Upon pulling this segment through the 7-branes one finds (see figure 8),
\be
\pmatrix{ P_m \cr Q_m \cr} = (I-M_m)  \pmatrix{ p_{(m)}\cr q_{(m)} \cr}
\ee

We will first consider the case in which none of the monodromies around 
any of the collections of coincident 7-branes has an eigenvalue equal to
one.  This excludes the $I_n$ series.  
In this case the matrix 
$I-M_m$ is invertible for all $m$.  It follows that the junction 
$J_{c,m} = P_m {\vec w}_{p,m} + Q_m {\vec w}_{q,m}$ (with no sum on $m$)
is a ``minimal'' proper junction,  in the sense that it can not be divided 
by any integer greater than $1$ and remain proper.   
If it could,  then a proper junction would 
be related by a Hanany-Witten 
transition to a segment of string junction with fractional charges,
i.e. a fraction of 
$ (p_{(m-1)}, q_{(m-1)})$ and $(p_{(m)}, q_{(m)})$ to the left and right
of the branch cut respectively, 
where $gcd( p_{(m-1)}, q_{(m-1)} ) = gcd(p_{(m)}, q_{(m)})= 1$.
However Hanany-Witten transitions do not relate proper and fractional
strings.  

The fact that ${\vec J}_{c,m} = P_m {\vec w}_{p,m} + Q_m {\vec w}_{q,m}$ is, 
for  each $m$, a minimal proper junction is the essential fact that allows 
one to choose a metric for which (\ref{agpropone}) and 
(\ref{agproptwo}) are satisfied.  
Consider what would happen if this were not the case.
For instance suppose that  
$gcd(P_1,P_2,\cdots, Q_1,Q_2,\cdots) =2$ and
${\vec J}_{c,1}$ is minimal but ${\vec J}_{c,2}$ is non-minimal until divided
by two.
In this case ${\vec N}_{\tilde p,\tilde q}/2$ does not satisfy 
(\ref{agpropone}) and (\ref{agproptwo}).  ${\vec J}_{c,2}/2$ would be
proper and correspond to a trivial element of the center of $G_2$,
but ${\vec J}_{c,1}/2$ would be improper and correspond to a non-trivial
element of the center of $G_1$.
In fact however,  ${\vec J}_{c,m}$ is a minimal proper junction
for all $m$,  and one can choose a metric such that $j_m$ and or $k_m$ are 
either +1 for all $m$ or $0$ for all $m$.  
Note that it is never true that both $j_m$ and $k_m$ are zero for all $m$
if $gcd(P_1,P_2,\cdots, Q_1,Q_2,\cdots) >1$,  since the fractional null
junctions are improper and represent non-trivial elements of the center
of the universal cover.  

We wish to emphasize that the correspondence
between the fractional null junctions and the trivially acting elements
in the center of the universal cover depends crucially on the choice of 
metric.  For instance,  the center of $G_m$ has an 
automorphism (complex conjugation) which flips the sign of $j_m$ or $k_m$. 
After acting with such an automorphism,
the null junctions correspond to different elements in the 
center of the universal cover which may not act trivially on 
physical states.    

We now consider the case in which there are $I_n$ fibers. 
These correspond to collections of 7-branes all of which
carry the same $(p,q)$ charge.  This case is somewhat more
subtle,  since $I-M_m$ is not invertible.  For example, if
the charge of the 7-branes is $(1,0)$,  then
a segment of string with charge $(1,0)$ passing under the
7-brane makes no contribution to the $P_m$ and $Q_m$ charge
upon pulling the string into the upper half plane;
\be
(I-M_m)\pmatrix{ 1 \cr 0\cr} = 0.
\ee
The only contribution to $P_m$ comes from the $q_{(m)}$ charge
of the string segment below the $I_n$ locus.
Consequently,  the junction ${\vec J}_{c,m}$ is not 
necessarily a minimal proper junction.  It is minimal
only after division by $q_{(m)}$. 

Let us make a choice of a generating pair ${\vec N}_{\tilde p,\tilde q}$
with $(\tilde p,\tilde q) = (0,1)$ or $(1,1)$.
The group associated with an $I_n$ fiber is $SU(n)$, with center
$Z_n$. 
If the vanishing cycle (7-brane charge) of this fiber is $(1,0)$, 
then the minimal proper junction is $n {\vec w}_{p,m}$.
Upon pulling the lower segment of the string loop into the upper 
half plane, one finds 
\be
{\vec J}_{c,m} = q_{(m)} n {\vec w}_{p,m}.
\ee
We can choose a metric such that 
\be
\tM({\vec J}_{c,m}) \cdot \tM( {\vec w}_{p,m})  =
q_{(m)} n \tM({\vec w}_{p,m}) \cdot \tM({\vec w}_{p,m})  =
2\pi i q_{(m)}. 
\ee
Then
\bea
\exp \left( i \frac{1}{gcd(P_1, P_2, \cdots, Q_1, Q_2 \cdots)}
		\tM({\vec J}_{c,m}) \cdot \tM({\vec w}_{p,m}) \right) = \nonumber \\
\exp \left( 2\pi i\frac{q_{(m)} }
		{gcd(P_1, P_2, \cdots, Q_1, Q_2 \cdots)} \right)
\eea
Now
\be
q_{(m)} = q_{(1)}+ Q_1 + Q_2 + \cdots + Q_m = 
\tilde q + Q_1 + Q_2 + \cdots + Q_m,
\ee
so that 
\bea
\exp \left( 2\pi i\frac{q_{(m)} }
		{gcd(P_1, P_2, \cdots, Q_1, Q_2 \cdots)} \right) = \nonumber \\
\exp \left( 2\pi i\frac{ \tilde q }
		{gcd(P_1, P_2, \cdots, Q_1, Q_2 \cdots)} \right) = \nonumber \\
\exp \left( 2\pi i \frac{1}{gcd(P_1, P_2, \cdots, Q_1, Q_2 \cdots)} \right)
\eea
This is precisely the same behavior as if ${\vec J}_{c,m}$ were
a minimal null junction;  there is no multiple of 
${\vec J}_{c,m}/gcd(P_1, P_2, \cdots, Q_1, Q_2 \cdots)$
which is both proper and ``smaller'' than ${\vec J}_{c,m}$. 

%%%%%%%%%%%%%%%%%%%%%%%%%%%%%%%%%%%%%%%%%%%%%%%%%%%%


\begin{thebibliography}{99}
  


\bibitem{AM}  P.~Aspinwall and D.~Morrison,
    {\it Nonsimply Connected Gauge Groups and Rational Points on
        Elliptic Curves}, JHEP {\bf 9807} (1998) 012, 
        hep-th/9805206 

\bibitem{FY} M.~Fukae, Y.~Yamada and S.~Yang,
        {\it Mordell-Weil Lattice via String Junctions}, Nucl.~Phys. {\bf B572} (2000) 71-94,
	hep-th/9909122. 

\bibitem{DZ} O.~DeWolfe and B.~Zwiebach, {\it String junctions for
    arbitrary Lie algebra representations} Nucl.~Phys. {\bf B541}
  (1999) 509-565, hep-th/9804210.


\bibitem{DW} O.~DeWolfe, {\it Affine Lie Algebras, String
    Junctions And 7-Branes}, Nucl. Phys. {\bf B550} (1999) 622-637,  
        hep-th/9809026.

\bibitem{DHIZ1} O.~DeWolfe, T.~Hauer, A.~Iqbal and B.~Zwiebach, {\it
    Uncovering Symmetries on [p.q] 7-branes: Beyond The Kodaira
    Classification}, hep-th/9812028.


\bibitem{DHIZ2} O.~DeWolfe, T.~Hauer, A.~Iqbal and B.~Zwiebach,
 {\it Uncovering Infinite Symmetries on $[p,q]$ 7-branes:
        Kac-Moody Algebras and Beyond.} hep-th/9812209.

\bibitem{DHIZ3} O.~DeWolfe, T.~Hauer, A.~Iqbal and B.~Zwiebach, {\it
    Constraints On The BPS Spectrum Of N=2, D=4 Theories With A-D-E
    Flavor Symmetry}, Nucl.~Phys. {\bf B534} (1998) 261-274,
    hep-th/9805220.

\bibitem{M-W} J.~Silverman and J.~Tate, {\it Rational Points on Elliptic Curves}, 
	Undergraduate Texts in Mathematics, Springer-Verlag, 1992.

\bibitem{NTY} M. Noguchi, S. Terashima, S. Yang, {\it N=2 Superconformal Field Theory
	with ADE Global Symmetry on a D3-Brane Probe}, Nucl.~Phys. {\bf B556} (1999) 115-151,
	hep-th/9903215.

\bibitem{SW} N.~Seiberg and E.~Witten, {\it Monopoles, Duality and Chiral Symmetry
	Breaking in N=2 supersymmetric QCD}, Nucl.~Phys. {\bf B431} (1994) 484-550,
	hep-th/9408099. 

\bibitem{OS} K.~Oguiso and T.~Shioda, {\it The Mordell-Weil Lattice of a Rational 
	Elliptic Surface}, Comment.~Math.~Univ.~St.~Pauli {\bf 40} (1991) 83.

\bibitem{HW} A. Hanany and E. Witten, {\it Type IIB Superstrings, BPS Monopoles, and Three 
	Dimensional Gauge Dynamics,} 
	Nucl.Phys. {\bf B492} (1997) 152-190, hep-th/9611230. 

\bibitem{Lerche} W. Lerche, {\it On the Heterotic/F-Theory Duality in
    	Eight Dimensions}, Progress in String Theory and M-Theory (Cargese 99), 
	hep-th/9910207.
  
\bibitem{Barrozo} M. C. Daflon Barrozo, {\it Map of Heterotic and Type
    IIB Moduli in 8 Dimensions},  Nucl.~Phys. {\bf B574} (2000) 189-218, 
	hep-th/9909178.
  
\bibitem{Imamura} Y. Imamura, {\it String Junctions and Their Duals
    in Heterotic String Theory}, Prog.Theor.Phys. {\bf 101} (1999)
  	1155-1164, hep-th/9901001.

\bibitem{US} A. Grassi,  Z. Guralnik and B. Ovrut, {\it  Junction Lattices, Calabi-Yau
	Three-folds, and Heterotic M-Theory,} in preparation.

\bibitem{W} E. Witten, {\it Phase Transitions in M Theory and F Theory,}
	Nucl.~Phys. {\bf B471} (1996) 195-216,  hep-th/9603150. 

\bibitem{IMS} K. Intriligator, D. Morrison and N. Seiberg, {\it Five dimensional 
	supersymmetric gauge theories and degenerations of Calabi-Yau Spaces,}
	Nucl.~Phys. {\bf B497} (1997) 56-100, hep-th/9702198. 

\bibitem{BIKMSV} M.~Bershadsky, K.~Intriligator, S.~Kachru, D.~Morrison,
        V.~Sadov, and C.~Vafa,{\it Geometric Singularities and Enhanced
        Gauge Symmetries},  Nucl.~Phys. {\bf B481} (1996) 215-252, 
        hep-th/9605200.

\bibitem{AG} P. Aspinwall and M. Gross, {\it The SO(32) Hetertotic String on
	a K3 Surface}, Phys.Lett.{\bf B387} (1996) 735, hep-th/9605131.

\bibitem{KV} S. Katz and C. Vafa, {\it Matter From Geometry,}
	Nucl.~Phys. {\bf B497} 146-154, hep-th/9606086.

\bibitem{AKM} P. Aspinwall, S. Katz, and D. Morrison, {\it Lie Groups, 
	Calabi-Yau Threefolds, and F Theory,} hep-th/0002012. 

\bibitem{K} K.~Kodaira, {\it On Compact Analytic Surfaces II}, Ann.~Math. {\bf 77}
	(1963) 563; {\it On Compact Analytic Surfaces III}, Ann.~Math. {\bf 78} 
	(1963) 1.

%%%\bibitem{AD} P.~Argyres and M.~Douglas,
%%%        {\it New Phenomena in $SU(3)$ Supersymmetric Gauge Theory},
%%%        Nucl.~Phys. {\bf B448} (1995) 93-126, hep-th/9505062.

  
\end{thebibliography}
\end{document}